\documentclass[superscriptaddress,aps,prd,nofootinbib,twocolumn,showpacs]{revtex4-1}

\usepackage[utf8]{inputenc}
\usepackage[british,UKenglish,USenglish,american]{babel}
\usepackage{amsmath}

\usepackage{graphicx}

\usepackage{float}

\usepackage{slashed,ulem}

\usepackage{mathrsfs}
\usepackage{amssymb}
\usepackage{bbm}

\usepackage{nicefrac}

\usepackage{microtype}

\usepackage{physics}

\usepackage{xcolor}

\usepackage{dsfont}

\usepackage{enumerate}

\newcommand\numberthis{\addtocounter{equation}{1}\tag{\theequation}}

\renewcommand{\bar}{\overline}

\usepackage{bm}

\newcommand{\e}{\mathrm{e}}

\newcommand{\comment}[1]{}

\begin{document}

\title{Quark matter in light neutron stars}

\author{Márcio Ferreira}
\email{marcio.ferreira@uc.pt}
\affiliation{CFisUC, Department of Physics, University of Coimbra, 
	P-3004 - 516  Coimbra, Portugal}

\author{Renan Câmara Pereira}
\email{renan.pereira@student.uc.pt}
\affiliation{CFisUC, Department of Physics, University of Coimbra, 
	P-3004 - 516  Coimbra, Portugal}

\author{Constança Providência}
\email{cp@uc.pt}
\affiliation{CFisUC, Department of Physics, University of Coimbra, 
	P-3004 - 516  Coimbra, Portugal}

\date{\today}

\begin{abstract}
Higher-order repulsive interactions are included in the three-flavor
NJL model  in order to describe the quark  phase of an hybrid star.
The effect of 4-quark and 8-quark  vector-isoscalar interactions
in the stability of hybrid star configurations is analyzed. 
The presence of a 8-quark vector-isoscalar channel is seen to be crucial
in generating large quark branches in the $M(R)$ diagram. 
This is due to  its  stiffening effect on the quark matter equation of state which
arises from the non-linear density dependence of the speed of sound.
This additional interaction channel allows for the appearance of a
quark core at moderately low NS masses, $\sim 1M_{\odot}$, and provides the required repulsion to preserve the star stability up to $\sim2.1M_{\odot}$. Furthermore, we show that both the heaviest NS mass generated, $M_{\text{max}}$, and its radii, $R_{\text{max}}$, are quite sensitive to the strength of 8-quark  vector-isoscalar channel, leading to a considerable decrease of $R_{\text{max}}$ as the coupling increases. This behavior imprints a considerable deviation from the 
purely hadronic matter equation of state in the $\Lambda(M)$ diagram,
which might be a possible signature of the quark matter existence,
even for moderately low NS masses, $\sim 1.4\, M_\odot$. 
The resulting $M(R)$ and $\Lambda(R)$ relations are in accordance with the latest astrophysical constraints from NICER and Ligo/VIRGO observations, respectively.  

\end{abstract}

\maketitle

\section{Introduction}
\label{introduction}

Neutron stars (NS) have been the focus of many experimental and theoretical studies in astrophysics, nuclear and particle physics. Their inner composition still remains an open question. The extreme densities reached in NS cores might originate some exotic matter, such as hyperons, Bose-Einstein condensates or quark matter \cite{Glendenning2012}.

The two solar mass pulsars PSR J1614-2230 ($M=1.908\pm$0.016
$M_\odot$) and PSR J0348+0432 ($M=2.01\pm$0.04 $M_\odot$)
\cite{Antoniadis:2013pzd}  and   MSP J0740+6620 \cite{Cromartie2019},
($M=2.14^{+0.10}_{-0.09} M_\odot$) impose tight constraints on the nuclear matter equation of state (EoS).
 Multi-messenger astrophysics that combines astrophysical observations
 of different type, electromagnetic radiation, gravitational waves
 (GW) and different types of particles provide deeper insights on NS properties. 
The analysis by the LIGO/Virgo collaborations of the GW from the NS merger GW170817 
gave us  important information about the NS structure
\cite{TheLIGOScientific:2017qsa,PhysRevLett.121.161101}, e.g., an
upper limit of the tidal deformability of a NS star, that allows us to
to set extra constraints on the high density EoS. Moreover, the
detection of the gamma-ray burst (GRB) GRB170817A \cite{grb}, and the
electromagnetic transient AT2017gfo \cite{kilo}  that followed up the
GW170817 event has further   established constraints on the lower limit of the tidal deformability \cite{Radice2017,Radice2018,Bauswein2019,Coughlin2018,Wang2018}.
The Neutron Star Interior Composition Explorer (NICER) experiment is
presently another important source of observational data that may shed
 some light into the structure of NS. Recently, two different teams of
 NICER have estimated the mass and radius of the millisecond-pulsar PSR J0030+0451 \cite{Riley_2019}.

While massive pulsars rule out soft EoS at high densities, a too stiff
EoS, which gives rise to  large radii, is incompatible with the tidal deformability from  GW observations \cite{Alford:2019oge}. 
The high density region of the EoS is thus severely constrained, which
may exclude exotic, i.e. non-nucleonic, degrees of freedom inside NS, such as quark matter. However, the existence of a first order phase transition from hadronic to quark matter, depending on its properties, may balance the two features mentioned above and still explain the observational data \cite{Alford:2019oge}. 
Detecting observational signatures that indicate the presence of exotic matter inside neutrons stars is a major difficulty.
For instance, it is hard to 
establish a clear physical distinction between a purely hadronic NS and one with a quark core solely from NS observables, such as the star mass, radius and tidal deformability.
However, the presence of a first order phase transition between hadronic and quark matter can lead to observational signatures that could be exploited in more neutron star binary mergers observations, favoring the hypothesis of quark matter in the neutron star core \cite{Most:2018eaw,Alford:2019oge,Weih:2019xvw}.

One way to study quark degrees of freedom in NS matter is through effective models, which incorporate the most important properties and symmetries of the strong interactions.
The NJL model is an widely used effective model of QCD. Some of its
applications are the study of the phase diagram of QCD, the behavior of
mesons at finite temperature and density and also to study of the
possible existence of quark matter inside neutron stars \cite{Schertler1999,Hanauske:2001nc,Baldo2002,Menezes2003,Pagliara:2007ph,Bonanno:2011ch,Lenzi:2012xz,Masuda:2012ed,Klahn:2013kga,Logoteta:2013ipa}.
The NJL model Lagrangian is built considering symmetry preserving interactions, specially chiral symmetry \cite{Hatsuda:1994pi,Buballa:2003qv}.

A possible approach to construct an hybrid EoS is the two model
approach: one that describes the hadronic (confined) phase and a
second model describing  the quark  (deconfined) phase. The matching
of the two EoS may be carried out within different approaches, in
particular considering local charge neutrality or global charge
neutrality \cite{Glendenning2000}. In the present approach we will consider a
Maxwell construction to describe a first-order phase transition from
hadron matter to a quark phase. This approach is considered to be
quite realistic if the surface tension between
hadron and quark matter, a still unknown quantity, is large.
This methodology has been widely used, where an hadronic model and an independent quark model were considered, see \cite{Pagliara:2007ph,Benic:2014iaa,Benic:2014jia,Zacchi:2015oma,Pereira:2016dfg,Wu:2018kww}. 
Using the NJL model to describe the quark phase of a hybrid EoS, previous works have successfully predicted neutron stars with at least $2M_\odot$ \cite{Bonanno:2011ch,Pereira:2016dfg}. 
The presence of the vector-isoscalar interaction was shown to be very important in stiffening the EoS to sustain $2M_\odot$. The inclusion of 8-quark interactions in the scalar and in the vector-isoscalar channel within the two-flavor NJL model was explored in \cite{Benic:2014iaa,Benic:2014jia} in the context of hybrid stars. In \cite{Ranea-Sandoval:2015ldr}, local and nonlocal NJL models with vector interaction among were seen to typically give no hybrid stars (or just small quark branches).

It has been shown by several authors that the onset of the $\Delta$
may compete with the onset of hyperons, and due to its large isospin
and the still lack of information to fix the coupling constants these
particles may set in at densities below the onset of hyperons, just
above saturation density
\cite{Drago2014,Ribes2019,Li_2019,Li:2019fqe}. In  particular, the
onset of $\Delta$s may occur in low mass  stars making compatible
relativistic mean-field models with the constraint set by GW170917 on
the tidal deformability. In the present work, we will show an
alternative scenario and will show that the onset of quarks at
densities below twice saturation density may also have a similar
effect of pushing down the tidal deformability of stars with masses
$\sim 1.4M_\odot$ or below.

Using a constant-sound-speed parametrization for the  
high-density EoS region \cite{Alford:2013aca,Alford:2015gna}, the
authors concluded that for a strong first-order phase transition to
quark matter to be  compatible with $M_{\text{max}}>2M_{\odot}$ requires a 
large speed of sound in the quark phase, $v_s^2\gtrsim 0.5$ for soft hadronic 
EoS and $v_s^2\gtrsim0.4$ for stiff hadronic EoS. Using the same formalism,
the work \cite{Han:2019bub} points in the same direction: strong
repulsive interactions in quark matter are required to support the NS
masses $M\gtrsim 2.0M_{\odot}$.

In \cite{Annala2019}, the authors studied the possibility of occurrence of
stars with  quark cores, imposing well known constraints, both
observational and theoretical  ab-initio calculations, to a large set of EoS built using
metamodels parametrized by the speed of sound. They propose that
1.4$M_\odot$ stars are compatible with hadronic stars. Besides, they
infer that  massive
stars with a mass $\approx 2M_\odot$ and a  speed of sound 
not far from  the conformal limit will have large quark cores. 
We would like to understand whether it is possible to arrive to similar
conclusions starting from a set of quark matter EoS that satisfy a given
number of constraints set by properties of mesons in  the vacuum which, also have been derived from a model with intrinsic chiral symmetry.

To attain this aim, we will work in the framework of the 
three-flavor NJL model, and we will analyze the effect of 4-quark and 8-quark vector-isoscalar interactions in hadron-quark hybrid EoS.
NJL models typically give rather low values for the speed of sound in
  the quark matter phase ($v_s^2\sim0.2-0.3$) and have a small
  dependence on the density. Furthermore, the speed of sound is quite
  insensitive to the NJL model parameters $\Lambda,\,m_{u,d},\,m_s,\,G_S,\,G_D$,
  i.e. the cutoff, current masses and couplings of the scalar and
  t'Hooft terms.
We will  investigate the impact of the vector interactions in the
speed-of-sound and in the quark phase and thus on the stability of hybrid stars sequences.
Moreover, exploring these additional interactions, we will analyze the possibility of having quark cores in light NS and, at the same time, fulfill all observational constraints.

This paper is organized as follows: in Section \ref{model_and_formalism} the quark model is detailed. 
The results are presented in Section \ref{results} followed by our conclusions, in Section \ref{conclusions}.

\section{Model and Formalism}
\label{model_and_formalism}

The SU$(3)_f$ NJL Lagrangian density, including four and six scalar-pseudoscalar interactions and four and eight vector-isoscalar interactions is:
\begin{align*}
\mathcal{L} & = 
\bar{\psi} 
\qty(
i\slashed{\partial} - \hat{m} + \hat{\mu} \gamma^0 
) 
\psi 
\\
& + G_S  \sum_{a=0}^8
\qty[ \qty(\bar{\psi} \lambda^a \psi)^2 + 
\qty(\bar{\psi} i \gamma^5 \lambda^a \psi)^2 ]
\\
& - G_D \qty[  
\det\qty( \bar{\psi} \qty(1+\gamma_5) \psi ) + 
\det\qty( \bar{\psi} \qty(1-\gamma_5) \psi )  ]
\\
& - G_\omega \qty[ (\bar{\psi}\gamma^\mu\lambda^0\psi)^2 + (\bar{\psi}\gamma^\mu\gamma_5\lambda^0\psi)^2 ]
\\
& - 
G_{\omega \omega} 
\qty[ 
(\bar{\psi}\gamma^\mu\lambda^0\psi)^2 + 
(\bar{\psi}\gamma^\mu\gamma_5\lambda^0\psi)^2 
]^2
.
\numberthis
\label{eq:SU3_NJL_lagrangian}
\end{align*} 
The diagonal matrices $\hat{m}=diag\qty(m_u, m_d, m_s )$ and $\hat{\mu}=diag\qty(\mu_u, \mu_d, \mu_s )$ are the quark current masses and chemical potential matrices, respectively. The matrices $\lambda^a$ with components $a=1,2...8$, are the Gell-Mann matrices of the SU(3) group while, the zero component, is a matrix proportional to the identity matrix, $\lambda^0=\sqrt{\nicefrac{2}{3}} \mathds{1}$. The quark field has $N_f$-components in flavor space.

The NJL model is nonrenormalizable in four dimensional space-time. Hence some regularization procedure must be employed in order to regularize the integrals. Alongside, the Matsubara formalism to derive the thermodynamical potential we are going to regularize the integrations using the 3-momentum cutoff regularization.

The multi-quark interactions considered are all chiral symmetry preserving. The four scalar and pseudoscalar quark interaction is present in the original formulation of the NJL model and is essential to incorporate in the model spontaneous chiral symmetry breaking. The 't Hooft determinant for three quark flavours corresponds to a six quark interaction which incorporates the explicit $U_A(1)$ symmetry breaking in the model. Incorporating vector interaction in the model has been found to be necessary to model the medium to high density behaviour of the EoS and predict $2M_\odot$ neutron stars. The inclusion of all possible chiral-symmetric set of eight quark vector interactions was performed in \cite{Morais:2017bmt} in order to study the masses of the lowest spin-0 and spin-1 meson states. Following previous works, the vector-isoscalar quark interactions have been showed to be essential to build $2M_\odot$ neutron stars.

In the present work, we will restrict our analysis to four and eight
vector-isoscalar quark interactions and study their  influence on the
EoS of hybrid neutron stars.  These vector interactions have free
coupling constants, $G_\omega$ and $G_{\omega \omega}$
respectively. In general, both of these couplings can be fixed in the
vacuum by fitting the omega meson mass. However, as discussed in  the
literature \cite{Buballa:2003qv,Fukushima:2008wg}, the vector-isoscalar 
terms are proportional to density degrees of freedom and their couplings might be
density dependent.
Hence, to take into account the possible in-medium dependence of the vector couplings $G_\omega$ and $G_{\omega \omega}$, we will not fix their magnitudes in the vacuum and leave them as free parameters. As in our previous works \cite{Pereira:2016dfg}, we will study different models defined by different values for the ratios $\xi_\omega = G_\omega / G_S$ and $\xi_{\omega \omega} = G_{\omega \omega} / G_S^4$.\\

The thermodynamical potential of the NJL model is calculated in the mean field approximation (MF), where the product between quark bilinear operators are linearized around their mean field values, and a linear Lagrangian density can be obtained (for more details on the linear product between $N$ operators see \cite{CamaraPereira:2020rtu}). The quark fields can then be integrated out.

Using the  Matsubara formalism and the linearized Lagrangian density, the MF thermodynamical potential of the NJL model, $\Omega$, is derived from the lagrangian written in Equation (\ref{eq:SU3_NJL_lagrangian}). For finite temperature and chemical potential it can be written as:
\begin{align*}
\Omega - \Omega_0 & = 
2 G_S  
\qty(  
\sigma_u^2 + \sigma_d^2 + \sigma_s^2 
) 
\\
& 
- 4 G_D \sigma_u\sigma_d \sigma_s 
\\
& 
- \frac{2}{3} G_\omega
\qty( 
\rho_u + \rho_d + \rho_s 
)^2  
\\
& 
-\frac{4}{3 }G_{\omega \omega}
\qty( 
\rho_u + \rho_d + \rho_s 
)^4
\\  
& - 2 T N_c \sum_{i=u,d,s} 
\int_0^{\Lambda} \frac{ \dd[3]{p} }{(2\pi)^3}   
\ln \qty( 1 + \e^{-(E_i+\tilde{\mu}_i)/T} ) 
\\  
& - 2 T N_c \sum_{i=u,d,s} 
\int_0^{\Lambda} \frac{ \dd[3]{p} }{(2\pi)^3}     
\ln \qty( 1 + \e^{-(E_i-\tilde{\mu}_i)/T}  ) 
\\  
& - 2 N_c \sum_{i=u,d,s} 
\int_0^{\Lambda} \frac{ \dd[3]{p} }{(2\pi)^3}    
E_i .
\numberthis
\label{NJLpot}
\end{align*}
The constant $\Omega_0$ is calculated in such a way that the potential vanishes in the vacuum. Also, $E_i=\sqrt{p^2+M_i^2}$ and $\sigma_i$ and $\rho_i$ are the condensate and density of the quarks with flavor $i$, respectively.

For $i\ne j\ne k\in \{u,d,s\}$, the effective mass, $M_i$, and effective chemical potentials, $\tilde{\mu}_i$, are found to be:
\begin{align*}
M_i = m_i 
& - 4G_S \sigma_i
+ 2G_D \sigma_j \sigma_k ,
\numberthis
\\
\tilde{\mu}_i = 
\mu_i 
& - 
\frac{4}{3} 
G_\omega 
\qty( 
\rho_i+\rho_j+\rho_k
)
\\
& - 
\frac{16}{9} 
G_{\omega \omega} 
\qty( 
\rho_i+\rho_j+\rho_k
)^3
.
\numberthis
\end{align*}

In the MF approximation the thermodynamical potential must be stationary with respect to the effective mass, $M_i$, and effective chemical potentials \cite{Buballa:2003qv}, $\tilde{\mu}_i$, i.e.,
\begin{align}
\pdv{\Omega}{M} = 
\pdv{\Omega}{\tilde{\mu}} = 0 .
\end{align}
Applying these stationary conditions to the thermodynamical potential yields a closed expression for the quark condensate, $\sigma_i$, and density, $\rho_i$. For the explicit expressions see \cite{Ferreira:2020evu}.

The quark sector of the cold hybrid EoS can be easily calculated from Equation (\ref{NJLpot}) in the $T=0$ limit. 
The pressure and energy density are given by
\begin{align}
P & = -\Omega , 
\label{def.pressao}
\\
\epsilon & = -P +  \sum_i \mu_i \rho_i. 
\label{def.energia}
\end{align}

Aside from the free vector couplings, $G_\omega$ and $G_{\omega \omega}$, the remaining parameters of the model are fixed in order to reproduce the values of some meson masses and decay constants. The used parameter set can be found in Table \ref{tab:2}. In Table \ref{tab:3} we present the values of some meson masses and leptonic decay constants within the parameter set in Table \ref{tab:2} and the respective experimental values.

\begin{table}[ht!]
\begin{tabular}{cccccccc}
    \hline
    \hline
$\Lambda$  & $m_{u,d}$ & $m_s$  & $G_S\Lambda^2 $ & $G_D\Lambda^5 $ & $M_{u,d}$ & $M_s$ \\
\text{[MeV]}      &  [MeV]    & [MeV]   &               &                  & [MeV] &  [MeV]\\
\hline
   \hline
623.58 & 5.70   & 136.60 & 1.67 &  13.67 & 332.2   & 510.7 \\
   \hline
\end{tabular}
\caption{Parameters of the NJL  model used in the present work: $\Lambda$ is the model cutoff, $m_{u,d}$ and $m_{s}$ are the quark current masses, $G_S$ and $G_D$ are coupling constants. $M_{u,d}$ and $M_{s}$ are the resulting constituent quark masses in the vacuum. This parameter set yields, in the vacuum, a light quark condensate of $\expval{\bar{q}_l q_l}^{1/3}=-243.9~\mathrm{MeV}$  and strange quark condensate of $\expval{\bar{q}_s q_s}^{1/3}=-262.9~\mathrm{MeV}$.}
\label{tab:2}
\end{table}

\begin{table}[ht!]
    \begin{tabular}{ccc}
    \hline
    \hline
     & NJL SU(3) {   }& Experimental \cite{Agashe:2014kda} \\
        \hline
          \hline
    $m_{\pi^{\pm}}$ [MeV]     & 139.6  & 139.6 \\
    $f_{\pi^{\pm}}$ [MeV]     & 92.0   & 92.2 \\
    $m_{K^{\pm}}$ [MeV]       & 493.7  & 493.7 \\
    $f_{K^{\pm}}$ [MeV]       & 96.4   & 110.4 \\
    $m_{\eta}$ [MeV]          & 515.6  & 547.9 \\
    $m_{\eta'}$ [MeV]         & 957.8  & 957.8 \\
    \hline
    \end{tabular}  
    \caption{The  masses and decay constants of several mesons within the model and the respective experimental values. }
  \label{tab:3}
\end{table}

The NJL model pressure and energy density are defined up to a constant $B$, analogous to the MIT bag constant \cite{Pagliara:2007ph}. It it essential in building hybrid EoS that sustain two-solar mass neutrons stars. In \cite{Pagliara:2007ph,Pereira:2016dfg}, $B$ was fixed by requiring that the deconfinement occurs at the same baryonic chemical potential as the chiral phase transition. More recently in \cite{Han:2019bub}, an effective bag constant was also used to control the density at which the phase transition from hadron to quark matter happened. In the presence of a finite bag constant, the quark EoS is modified by $P \to P + B$ and $\epsilon \to \epsilon-B$. Hence the NJL quark EoS will be defined by three parameters: the model vector coupling ratios, $\xi_{\omega}= G_{\omega} / G_S$ and 
$\xi_{\omega\omega}= G_{\omega\omega} / G_S^4$ and the bag constant $B$.

For the hadronic part of the hybrid stars we use the DDME2 model
\cite{ddme2}. This is a relativistic mean-field model with density
dependent couplings   that describes two solar mass stars and  satisfies a well established set of nuclear
matter and finite nuclei constraints \cite{Dutra2014,Fortin2016}, including
the constraints set by the ab-initio calculations for neutron matter
using a chiral effective field theoretical approach
\cite{Hebeler2013}. This has been the low density constraint set in \cite{Annala2020}.

\section{Results}
\label{results}

Herein, we analyze the effect of the vector-isoscalar couplings 
$\xi_{\omega}= G_{\omega} / G_S$ and 
$\xi_{\omega\omega}= G_{\omega\omega} / G_S^4$ on the hybrid EoS and respective NS properties. 
The effect of the bag constant $B$ was already studied in \cite{Hanauske:2001nc,Klahn:2006iw,Pagliara:2007ph,Bonanno:2011ch,Lenzi:2012xz,Masuda:2012ed,Klahn:2013kga,Logoteta:2013ipa,Menezes:2014aka,Klahn:2015mfa,Pereira:2016dfg,Ferreira:2020evu}, where it was found that
the onset of quark matter in the hybrid EoS happens at lower densities as $B$ increases.
Although we have explored several values for $B$, we have decided to keep it
fixed  in the following analysis to $B=10$ MeV/fm$^{3}$.  
As free parameters, we consider 
$\{\xi_{\omega},\,\xi_{\omega\omega}\}$ which give a considerable
flexibility to span a wide range of EoS with the required properties. 
In the following, charge-neutral neutron star matter in
$\beta-$equilibrium, with  a first-order phase transition (via a
Maxwell construction) from hadronic matter to quark matter happens, is
studied.

\begin{figure}[!htb]
	\centering
	\includegraphics[width=1\columnwidth]{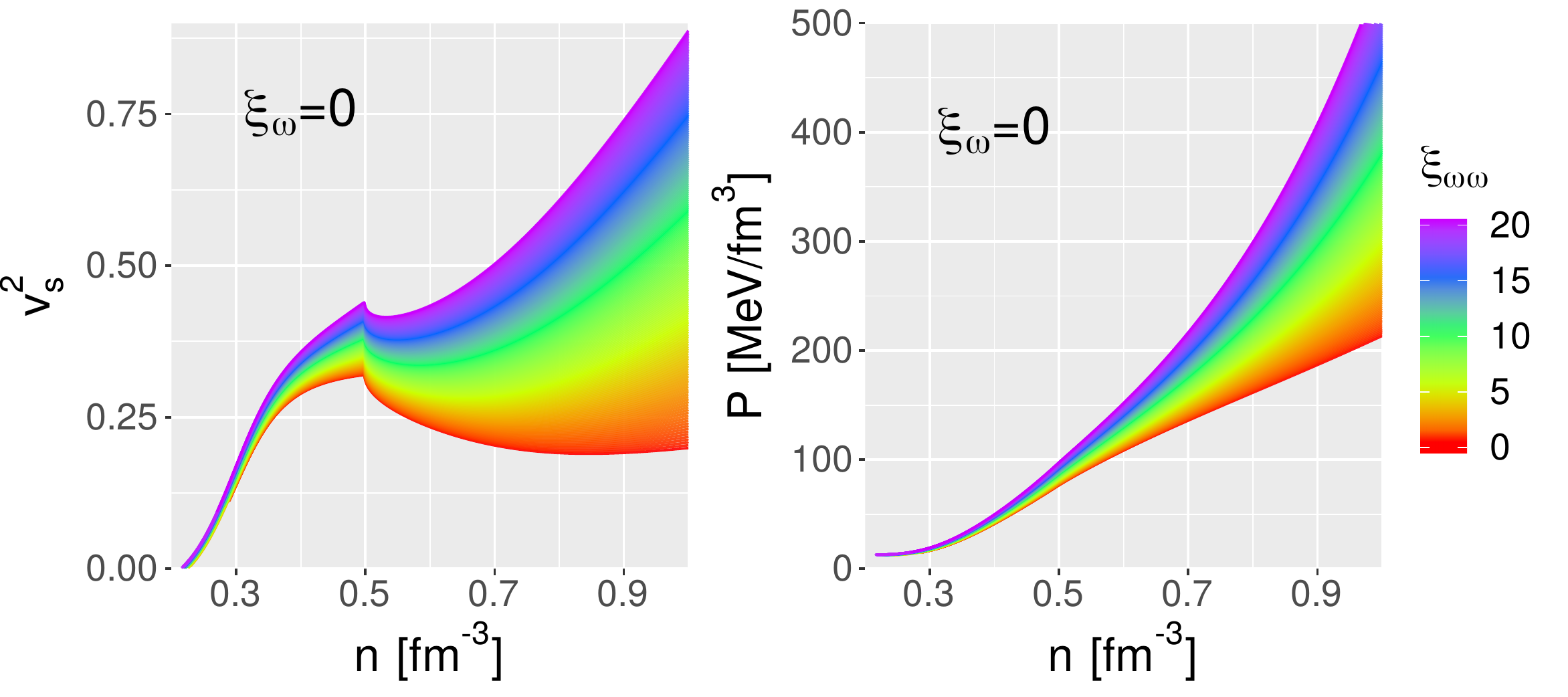}
	\caption{The speed of sound (left) and pressure (right) as a function of density for quark matter with $\xi_\omega=0.0$.  
		The color scale indicates the $\xi_{\omega\omega}$ value.}
\label{fig:vs}
\end{figure}

The main effect of the 4-quark vector term is to stiffen the quark
EoS and shift the onset of quark matter to larger densities as
discussed in \cite{Logoteta:2013ipa,Pereira:2016dfg}. Moreover, the larger the coupling
constant, $\xi_\omega$ the smaller the quark core. This behavior has
been described considering a  constant speed of sound model for
the quark phase \cite{Alford2015}.

\begin{figure}[!htb]
	\centering
	\includegraphics[width=0.8\columnwidth]{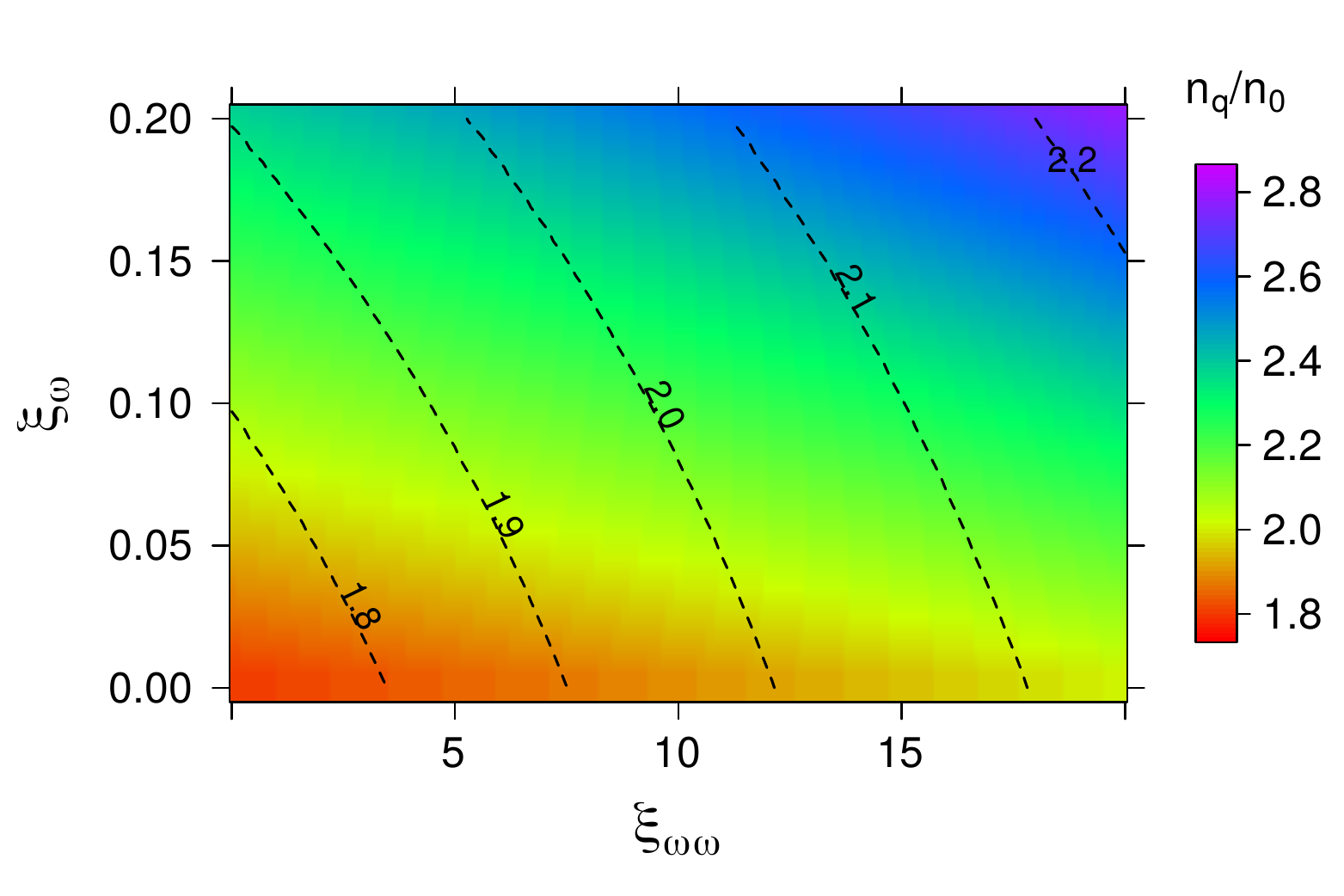}
	\caption{Onset density of quark matter, $n_{q}$ (in units of saturation density, $n_0=0.155$ fm$^{-3}$), as a function of both $\xi_{\omega}$
	and $\xi_{\omega\omega}$. The dashed lines represent the value of the maximum NS mass [in $M_{\odot}$] reached by each hybrid EoS, defined by $(\xi_{\omega},\xi_{\omega\omega})$. }
\label{fig:onset}
\end{figure}

Let us now analyze how $\xi_{\omega\omega}$ affects the quark matter
EoS.  Figure \ref{fig:vs} shows the pressure (right) and the speed of
sound squared (left) as a function of baryonic density for $\xi_{\omega}=0$ (herein, we use $c=1$).
The speed of sound, $v_s^2=dp/d\epsilon$,  characterizes how 
stiff the EoS is. It  is clear from both panels that  the
8-quark term, characterized by the coupling $\xi_{\omega\omega}$,
allows the quark EoS to  become stiffer so that  a larger quark core
will be sustained in the hybrid NS: this term gives rise
to a density dependent speed of sound that increases non-linearly with
density.
The main role of  $\xi_{\omega\omega}$ is played at large densities:
it affects in a much smaller extension the onset of quark
matter than the   $\xi_{\omega}$ coupling. 
This is clearly seen in Figure \ref{fig:onset}, where the 
onset density of quark matter, for each hybrid EoS, is shown  by a color degrade  in terms
of the parameters $\xi_{\omega\omega}$  and $\xi_{\omega}$. The change
of color is only slightly dependent on   $\xi_{\omega\omega}$.

The sudden decrease of  the speed of sound $v_s^2$ at $n\approx 0.5$
fm$^{-3}$ is due to the onset of  strangeness. Note, however,
that the appearance of the strange quark occurs via a crossover and thus
in a continuous way.  Since the vector terms introduced
are flavor invariant \cite{Buballa:2003qv}, the onset of strangeness does not depend of the
vector terms and is completely defined by the properties of the model
shown in Table \ref{tab:2}.  The amount of strangeness inside the
star, will, therefore, be determined by  the central density that
depends on both vector terms.

\begin{figure}[!htb]
	\centering
	\includegraphics[width=1\columnwidth]{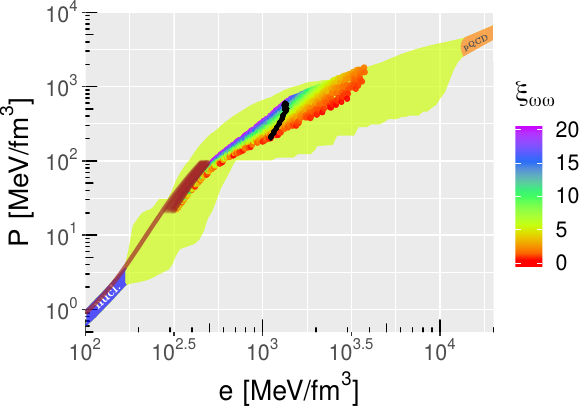}
	\caption{The EoS used in the present study in pressure vs
          energy density. The color scale refers to the parameter
          $\xi_{\omega\omega}$. At low densities  the DDME2 EoS is
          represented  followed by the hadron-quark phase transition
          at constant pressure (Maxwell construction). All EoS shown
          are causal. On the background  the contours of the region
          defined in \cite{Annala2020} for the acceptable EoS that
          interpolate between the neutron matter EoS  determined for
          a chiral effective  field theory approach in
          \cite{Hebeler2013}  and the pQCD EoS calculated in
          \cite{Kurkela2009}. The black dots identify the maximum mass  stars.
	}
\label{fig:annala}
\end{figure}

We plot in Fig. \ref{fig:annala} our set of EoS  on a pressure  vs
energy density graph for $\xi_{\omega}=0$, and include in 
the background the acceptable region of EoS
defined  in \cite{Annala2020}.  We conclude that our set of EoS covers
a quite large fraction of the proposed region. The red color indicates
a region with a speed of sound $v_s^2\lesssim 0.3$ as shown in
Fig. \ref{fig:vs}. Our most massive stars (purple
color)  lie close to the boarder of the region and  are associated
with central speed of sound well above  the conformal limit, which can
be as large as 0.9c.   Some interesting conclusions are: a) our set of
EoS also defines a change of slope. This could be due to the fact that
we work with a model with chiral symmetry incorporated. This kind of knee is also present
in other studies \cite{Tews2018}; b) we get low mass stars with a
quark core below the knee; c) our heaviest stars with a large quark core have a speed of sound
far from the conformal limit; d) the red dots identify EoS with a
speed of sound close to the conformal limit and lie in the center of
the region as obtained  in \cite{Annala2020}; 
e) the vector interactions considered in this work do not span
  the whole region of the Fig. \ref{fig:annala}. Including extra four
  and eight quark vector interactions, for instance in the scalar and vector-isovector channels, may increase this region. This is left as future work.

\begin{figure}[!htb]
	\centering
	\includegraphics[width=1\columnwidth]{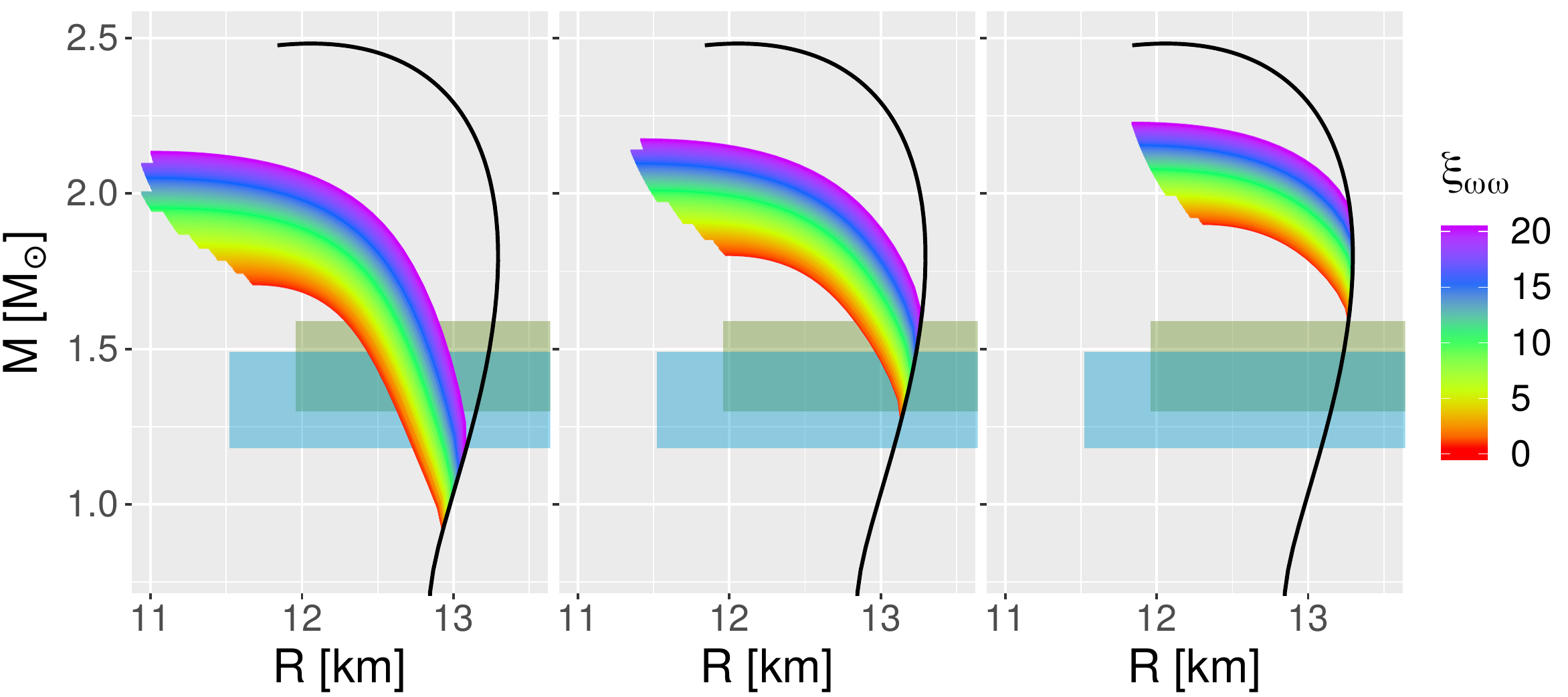}
	\caption{$M(R)$ diagrams for $\xi_\omega=0$ (left), $\xi_\omega=0.1$ (center), and 
	$\xi_\omega=0.2$ (right). The color scale indicates the $\xi_{\omega\omega}$ value and 
    the black line represents the purely hadronic sequence.
	The bag constant is fixed at $B=10$ MeV/fm$^{3}$. 
	The colored regions indicate the $(M,R)$ constraints obtained by two independent analysis using 
	the NICER x-ray data from the millisecond pulsar PSR J0030+0451 \cite{Riley_2019,Miller:2019cac}.}
	\label{fig:TOV}
\end{figure}

In order to study the  NS properties we have integrated the Tolmann-Oppeheimer-Volkof (TOV)
equations \cite{TOV1,TOV2} and the tidal deformabilities $\Lambda$ are calculated
as in \cite{Hinderer2008}.
Fig. \ref{fig:TOV} shows the $M(R)$ diagram for each hybrid EoS, parametrized by $(\xi_{\omega},\xi_{\omega\omega})$. For the sake of clarity, we have fixed $\xi_{\omega}$ in each panel:
$\xi_{\omega}=0.0$ (left), $\xi_{\omega}=0.1$ (center), and $\xi_{\omega}=0.2$ (right). The color scale encodes the value of $\xi_{\omega\omega}$. 
The effect of $\xi_{\omega}$ is clear: as its value increases, quarks
appear at  larger masses,   shorter quark star branches are obtained, which reach higher $M_{\text{max}}$.
As expected, given that both represent repulsive interactions,
$\xi_{\omega\omega}$ shows the same trend as $\xi_{\omega}$. Higher
values of $\xi_{\omega\omega}$ originate longer quark branches capable
of reproducing  more massive  NS.
The most interesting cases occur for smaller values of $\xi_{\omega}$ and for considerable values of $\xi_{\omega\omega}$, see left and center panels. 
Under these conditions, 
quarks are already present inside light NS, $M>0.9M_{\odot}$,
and it is still  possible to attain  quite massive and  compact NS, $M\approx2.2M_{\odot}$ and $R\approx 11$ km. 
For $\xi_{\omega\omega}>10$, hybrid NS with $M>1.9M_{\odot}$ that
predict already some quark content for $M\approx1.0M_{\odot}$ NS are possible.

We have represented two shaded regions in Figure \ref{fig:TOV} that indicate the $(M,R)$ constraints obtained by two independent analysis using the NICER x-ray data from the millisecond pulsar PSR J0030+0451 \cite{Riley_2019,Miller:2019cac}. The set of hybrid EoS in the present work are in good agreement with both constraints.

\begin{figure}[!htb]
	\centering
	\includegraphics[width=1\columnwidth]{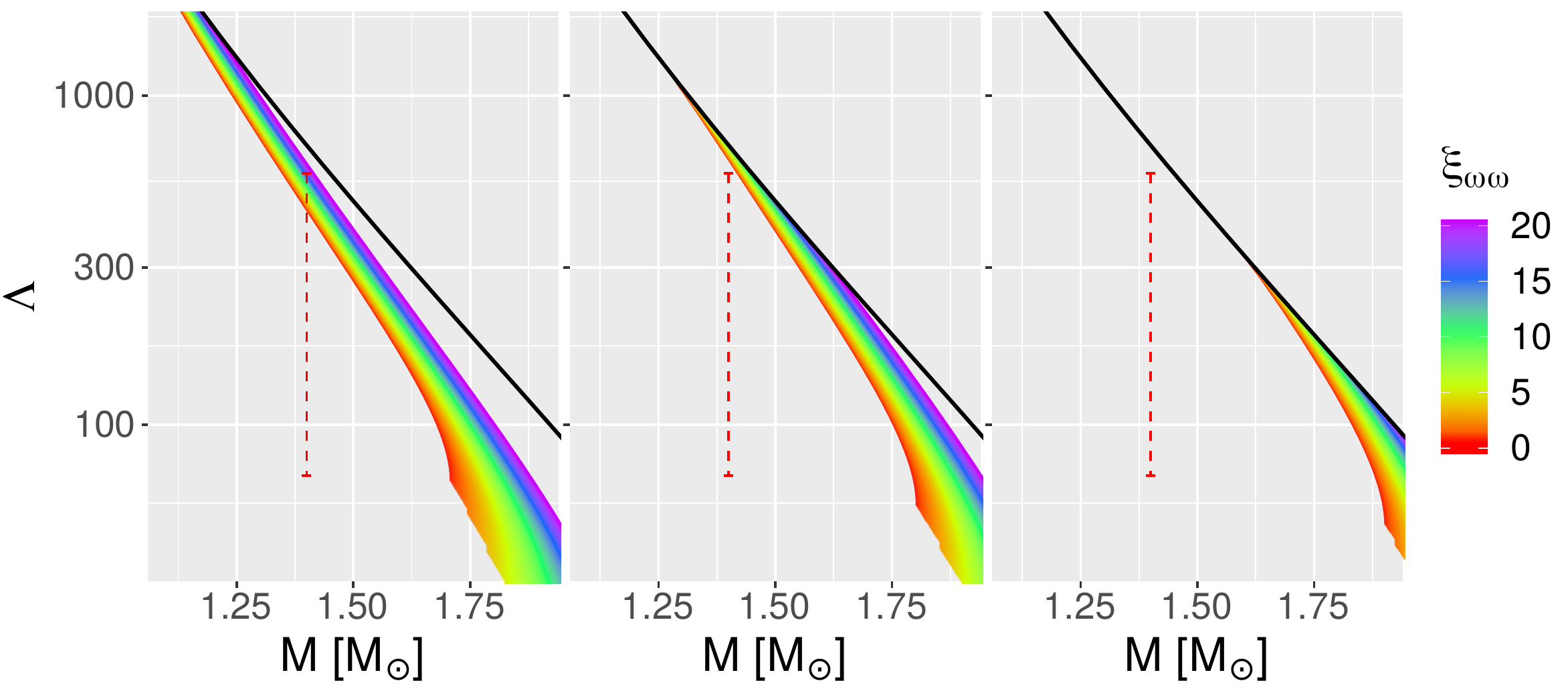}
	\caption{$\Lambda(M)$ diagrams for $\xi_\omega=0$ (left), $\xi_\omega=0.1$ (center), and $\xi_\omega=0.2$ (right). 
	The color scale indicates the $\xi_{\omega\omega}$ value and the black line represents the purely hadronic sequence.
	The bag constant is fixed at $B=10$ MeV/fm$^{3}$. 
	The dashed red line indicates the constraint $70<\Lambda_{1.4M_{\odot}}<580$ (90\% level) from the  GW170817 event \cite{Abbott18}}
\label{fig:lambda}
\end{figure}

The $\Lambda(R)$ diagrams are shown in Figure \ref{fig:lambda}. 
Like in Figure \ref{fig:TOV}, we show three panels:
$\xi_{\omega}=0.0$ (left), $\xi_{\omega}=0.1$ (center), and $\xi_{\omega}=0.2$ (right).
The red dashed line represents the constraint $70<\Lambda_{1.4M_{\odot}}<580$ (90\% level) obtained
from the  GW170817 event \cite{Abbott18}. We see that, with the combination of low $\xi_{\omega}$ 
and high $\xi_{\omega\omega}$, it is possible to generate an hybrid
EoS  that softens the hadronic EoS (solid black line) at low baryonic
densities, and satisfies the GW170817 $\Lambda_{1.4M_{\odot}}$ constraint.
Another interesting result is that the radius of the heaviest stable NS, $R_{\text{max}}$, is quite sensitive to 
the $\xi_{\omega\omega}$ value, and it is possible to predict sequences in the $\Lambda(R)$ diagram that clearly deviate from the purely hadronic EoS one. 
Small  values  of $\Lambda$ for a low/intermediate mass star could be
an important signature indicating the presence of quark matter in NS, which would be accessible through observational results on $(M_i,R_i,\Lambda_i)$.

\begin{figure}[!htb]
	\centering
	\includegraphics[width=0.9\columnwidth]{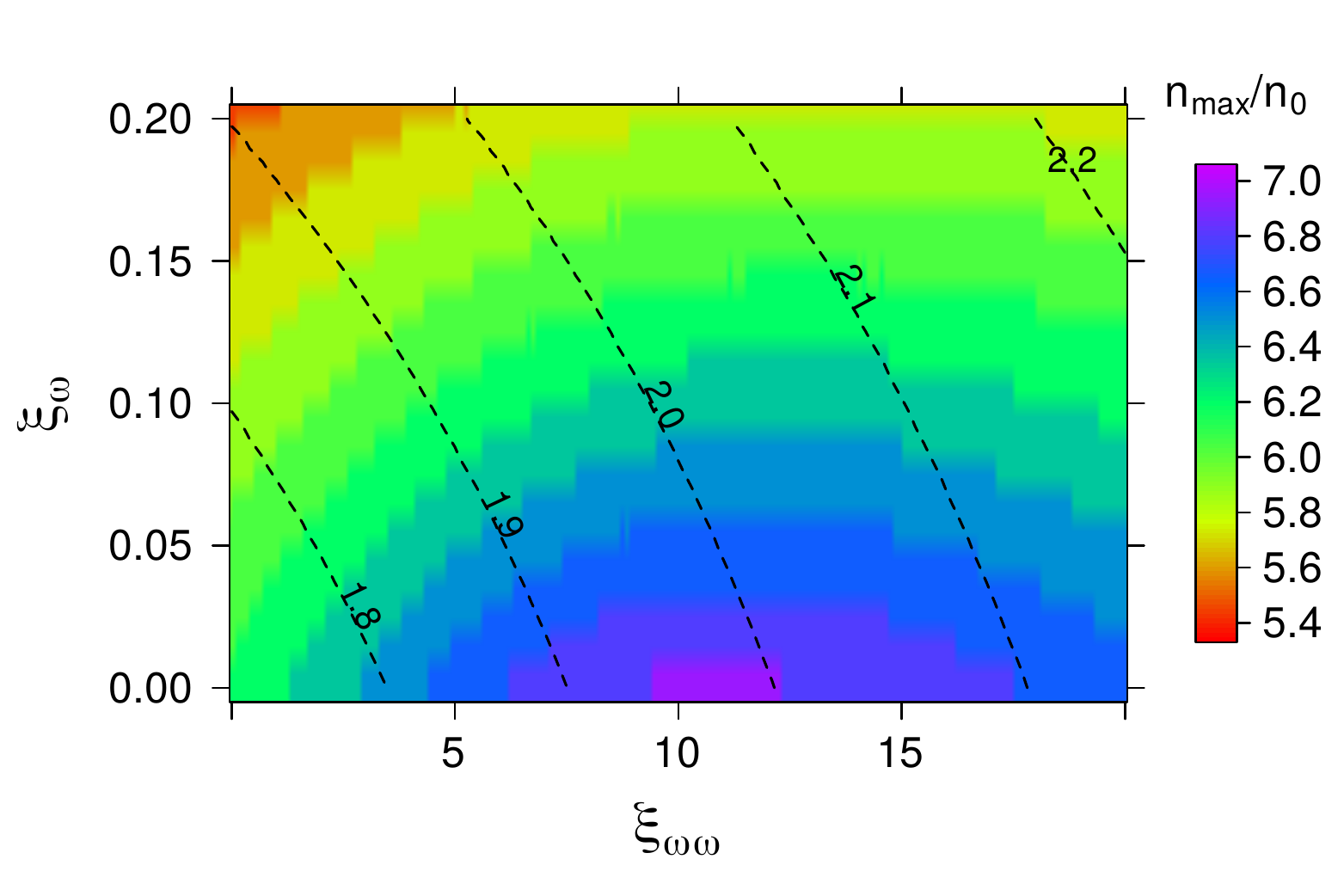}
	\caption{Central density at the maximum NS mass, $n_{max}$ [in units of saturation density, $n_0=0.155$ fm$^{-3}$], as a function of both $\xi_{\omega}$
	and $\xi_{\omega\omega}$. The dashed lines represent the value of the maximum NS mass [in $M_{\odot}$] reached by each hybrid EoS, defined by $(\xi_{\omega},\xi_{\omega\omega})$. }
\label{fig:nova}
\end{figure}

In Figure \ref{fig:nova}, we show how the central density, $n_{max}$ at the maximum NS, $M_{max}$, depends on $(\xi_{\omega},\xi_{\omega\omega})$. The overall effect of 
$\xi_{\omega}$ is to decrease the central density of $M_{max}$, while $\xi_{\omega\omega}$ shows a clear non-monotonic impact on $n_{max}$. The maximum value of $n_{max}$ is reached 
for $\xi_{\omega}=0$ and  $\xi_{\omega\omega}\approx 11$. 
This is already seen in Figure \ref{fig:TOV} (left panel), where the
$R_{max}$ shows a non-monotonic behavior: it increases up to
$\xi_{\omega\omega}=10$ and then starts to decrease for higher
$\xi_{\omega\omega}$ values.   Since the
onset of the s-quark occurs at $\approx 0.5$ fm$^{-3}$ independently
of the vector interaction, as we have seen before, 
we conclude that all stars have some fraction of
s-quarks. However, if $\xi_\omega> 0.1$ the amount of strangeness is
quite small. This behavior has also been found in hadronic matter with
hyperons: if the coupling to the vector mesons is strong the
strangeness content of the star is  small
\cite{Weissenborn13,Oertel2015}. It is interesting, however, to
realize that the 8-quark term stiffens the  EoS but still allows very
large central baryonic densities, and, as a consequence, a large
strangeness content.

\begin{figure}[!htb]
	\centering
	\includegraphics[width=0.9\columnwidth]{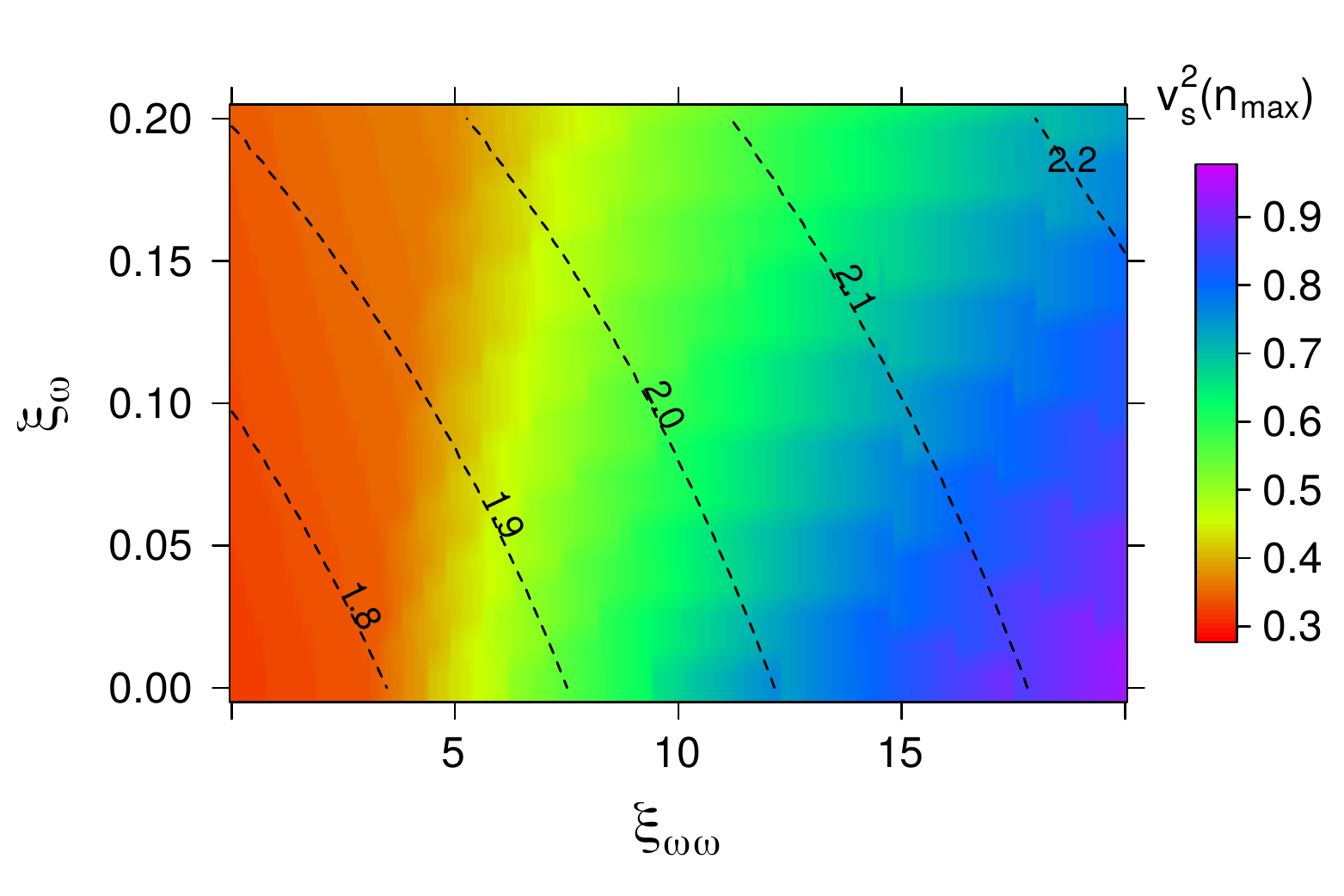}
	\caption{Speed of sound at the central density of the most massive stable NS, $v_s^2(n_{\text{max}})$, as a function of both $\xi_{\omega}$	and $\xi_{\omega\omega}$. The dashed lines represent the value of the maximum NS mass [in $M_{\odot}$] reached by each hybrid EoS, defined by $(\xi_{\omega},\xi_{\omega\omega})$.}
\label{fig:5}
\end{figure}

\begin{figure}[!htb]
	\centering
	\includegraphics[width=0.48\columnwidth]{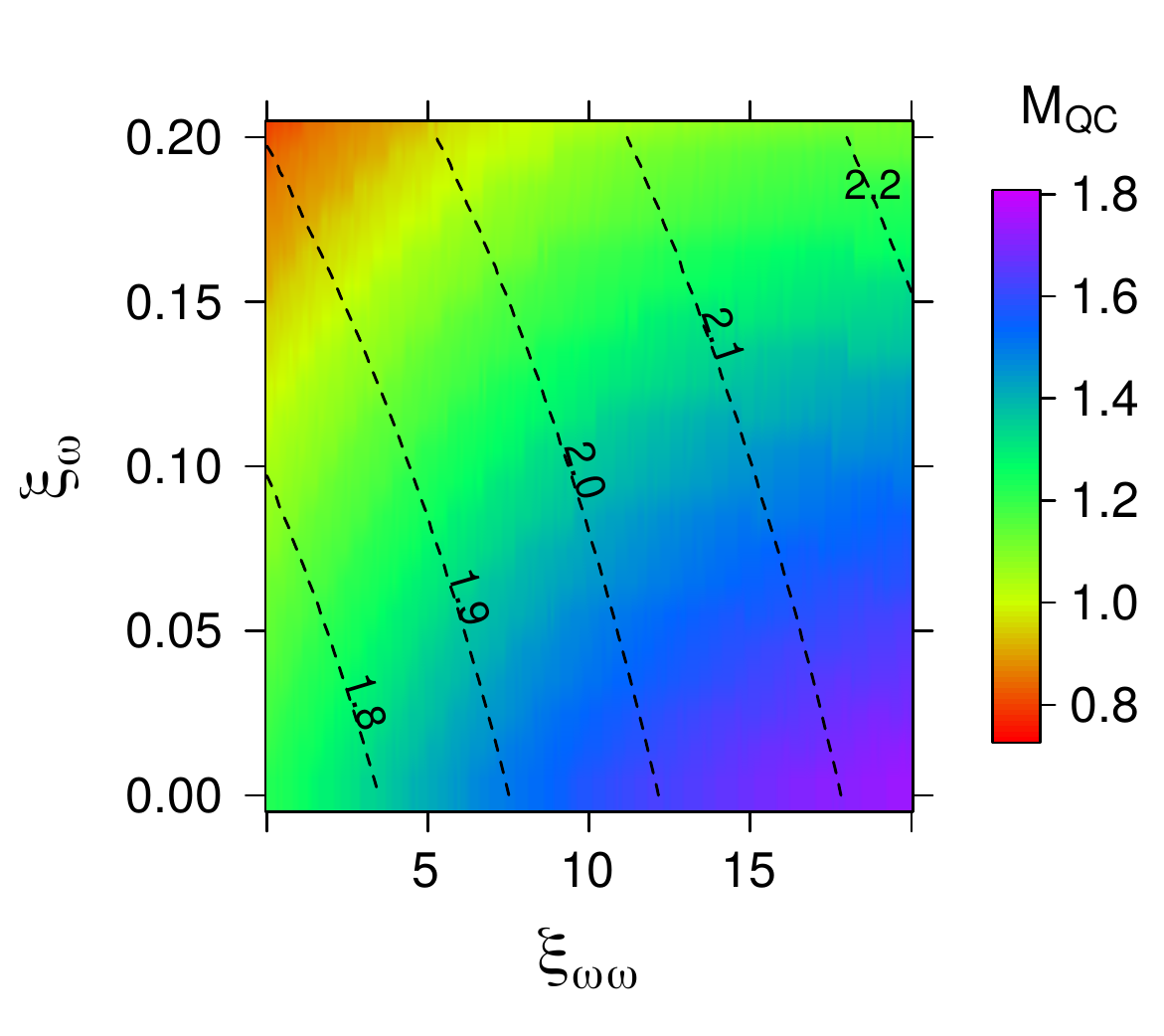}
	\includegraphics[width=0.48\columnwidth]{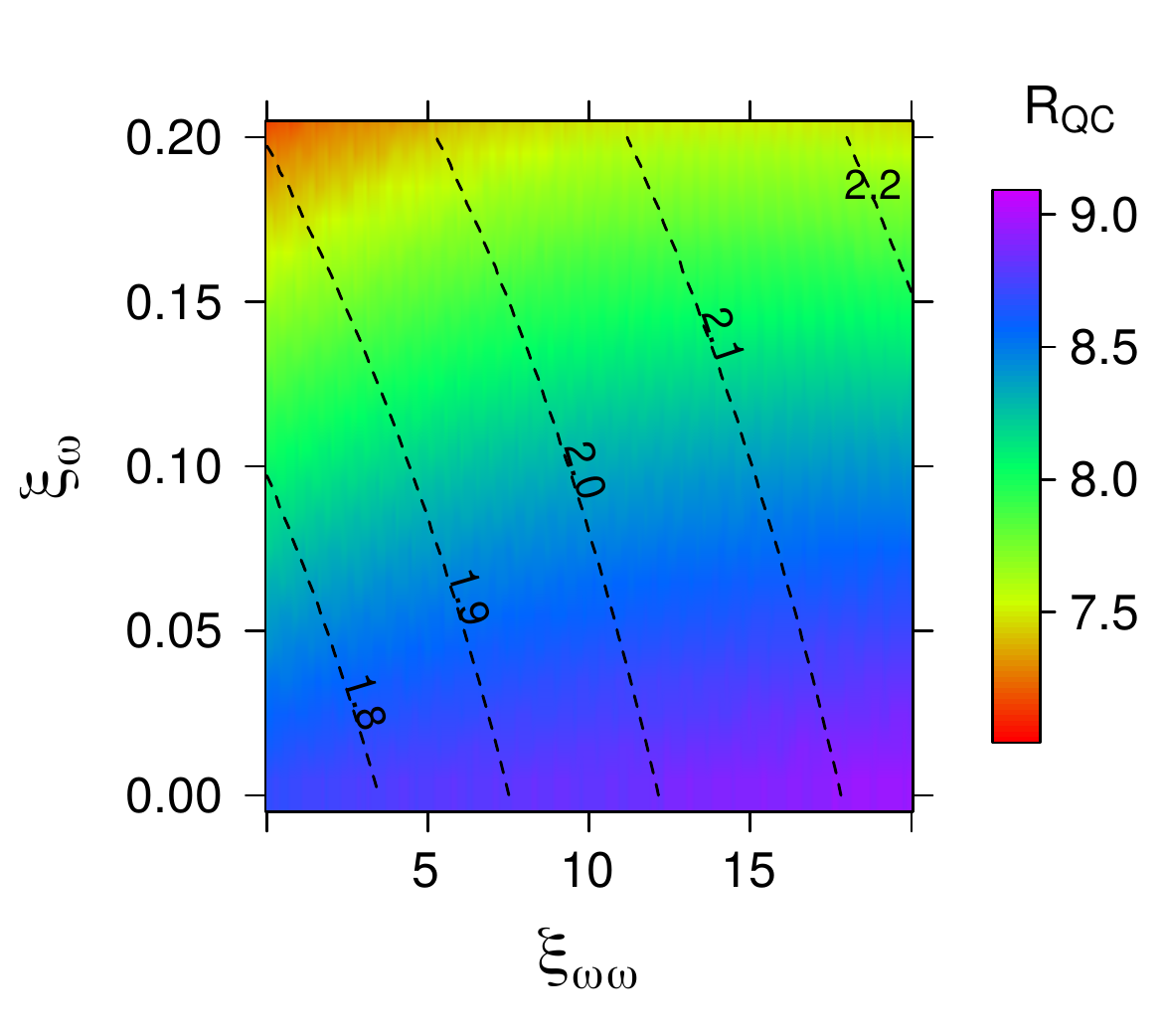}
	\caption{The quark core mass $M_{QC}$ [in $M_{\odot}$] (left)  and  radii $R_{QC}$ [in km] (right) as a function of both $\xi_{\omega}$
	and $\xi_{\omega\omega}$. The dashed lines represent the value of the maximum NS mass [in $M_{\odot}$] reached by each hybrid EoS, defined by $(\xi_{\omega},\xi_{\omega\omega})$. }
\label{fig:4}
\end{figure}

In Figure \ref{fig:5}, we display the speed of sound squared, $v_s^2$,
attained at the central density of the heavier NS ($M_{\text{max}}$)
for each hybrid EoS, i.e., $v_s^2(n_{\text{max}})$, which is a
function of  $(\xi_{\omega},\xi_{\omega\omega})$. $v_s^2$ is very
sensitive to  $\xi_{\omega\omega}$ and is only slightly  affected by $\xi_{\omega}$ .
To reach  massive NS cores, it is crucial to have large $v_s^2$ values. 
The quark core of $M\approx1.8M_{\odot}$ in Figure \ref{fig:4}, is
possible only  because the star has  
a very stiff quark matter phase, with $v_s^2\approx0.93$. \\

Let us now analyze how the quark core size depends on $(\xi_{\omega},\xi_{\omega\omega})$.
Figure \ref{fig:4} displays both the mass of the quark core, $M_{QC}$ (right panel), 
and the radii, $R_{QC}$ (left panel), as a function of $(\xi_{\omega},\xi_{\omega\omega})$.
We further indicate the maximum mass reached by each hybrid stars
through contour lines as before (black dashed lines).
For a fixed $\xi_{\omega}$ value, $M_{QC}$ increases with $\xi_{\omega\omega}$, reaching
a heavier quark core for low $\xi_{\omega}$ and high
$\xi_{\omega\omega}$. This is precisely when the central density is
the largest.  On the other hand,
for a fixed $\xi_{\omega\omega}$ value, $M_{QC}$ decreases as the value of $\xi_{\omega}$ gets bigger.
Therefore, the extremes of $M_{QC}(\xi_{\omega},\xi_{\omega\omega})$ lie in opposite regions:
the lighter quark core, $M\approx0.8M_{\odot}$, is found for $(\xi_{\omega}=0.2,\xi_{\omega\omega}=0)$ while the heavier, $M\approx1.8M_{\odot}$, is generated for $(\xi_{\omega}=0,\xi_{\omega\omega}=20)$. 
Actually, a quark core of $M\approx1.8M_{\odot}$ is generated in a region where $M_{\text{max}}\approx 2.1 M_{\odot}$, showing that 85\% of the star has quark degrees of freedom.  
Even though $R_{QC}$ displays a similar trend as $M_{QC}$, there is a greater sensitivity 
to $\xi_{\omega}$ than $\xi_{\omega\omega}$. 
Even for low $\xi_{\omega\omega}$ values,
the quark core radii can reach values as high as $9$ km, although two
solar mass stars are not attained for these values.
The contour lines representing $M_{\text{max}}$ reflect a much
stronger dependence on $\xi_{\omega\omega}$ than on $\xi_{\omega}$.

\section{Conclusions}
\label{conclusions}

In this work we have analyzed the effect of 4-quark and 8-quark vector-isoscalar interactions in hadron-quark hybrid EoS within the three flavor NJL model. 
Each hybrid EoS consists of charge-neutral matter in $\beta-$equilibrium, in which 
a first-order phase transition from hadronic to quark matter is present. 
We have analyzed how the stability of hybrid stars sequences and their
properties depend on the four and eight 
vector-isoscalar couplings, $\xi_{\omega}= G_{\omega} / G_S$ and 
$\xi_{\omega\omega}= G_{\omega\omega} / G_S^4$.

From the density dependence of the speed of sound of quark matter, one
clearly recognizes the stiffening effect of both interactions. This
behavior imprints interesting features  in the sequences of stable
star in the $M(R)$ diagram. We show that the size of the quark star
branch is quite sensitive to both couplings, particularly to the $\xi_{\omega\omega}$ coupling.
With a small value for $\xi_{\omega}$, there is a range of $\xi_{\omega\omega}$ values that 
predict quark matter in light NS, $\sim 1M_{\odot}$, and, at the same
time, are able to sustain a quark core in quite massive NS, i.e.,
$\sim2.1M_{\odot}$. 
Furthermore, the radii of the heaviest stable NS,
$R_{\text{\text{max}}}$, is highly dependent on the strength of
$\xi_{\omega\omega}$, leading to a considerable decrease of
$R_{\text{max}}$ as the coupling increases. As a consequence, for  a hybrid EoS a
considerable deviation from   the purely hadronic matter EoS
prediction for the  tidal deformability  $\Lambda(M)$ is obtained.
This occurs  even for moderate NS masses, $\sim 1.4\, M_\odot$, in accordance with the astrophysical constraints from NICER and LIGO/Virgo observations.

We have also discussed how the size of the quark core depends on $\xi_{\omega}$ and $\xi_{\omega\omega}$.
We have concluded that, for a fixed $\xi_{\omega}$ value, $M_{QC}$ increases with $\xi_{\omega\omega}$. While lighter quark cores, $\sim 0.8M_{\odot}$, are predicted for $(\xi_{\omega}=0.2,\xi_{\omega\omega}=0)$, the heaviest cores, $\sim 1.8M_{\odot}$, are generated in the opposite regime, i.e., $(\xi_{\omega}=0,\xi_{\omega\omega}=20)$. 
Quite massive quark cores, $\sim1.8M_{\odot}$, are predicted for hybrid EoS in each $M_{\text{max}}\approx 2.1 M_{\odot}$, showing that there are quark degrees of freedom in  85\% of the star.

Concerning the conclusions drawn in \cite{Annala2020}, we obtain
some similar results, in particular, we are able to describe two solar
mass stars with a central speed of sound squared below $0.4$, but more massive
stars require larger central values for the speed of sound.  However,
some other aspects in our study differ  from the ones discussed in  \cite{Annala2020}.
 We have obtained low mass stars with a quark core,   and we can describe very massive stars
with large quark cores and a speed of sound far from the conformal
limit. This is also in divergence with the conclusions drawn in
\cite{Alford:2019oge} because we  were able of getting large quark
cores even with a high central speed of sound, and the reason is that
 the model used to
perform our study allows for a density dependent speed of sound, with
a non-linear density dependence.

A low mass NS with a quark core would be confirmed if  together with
the  BNS tidal deformability and mass, also the dominant post-merger GW
frequency f peak would be measured. In \cite{Bauswein2018} it was
shown  that this frequency would  identify a first-order phase
transition. In the presence of a first order phase transition the f
peak comes at a much larger frequency: the larger the baryonic density
gap  at the phase transition the larger the frequency.

\section{Acknowledgments}

This work was partially supported by national funds from FCT (Fundação para a Ciência e a Tecnologia, I.P, Portugal) under the IDPASC Ph.D. program (International Doctorate Network in Particle Physics, Astrophysics and Cosmology), with the Grant No. PD/\-BD/128234/\-2016 (R.C.P.), under the Projects No. UID/\-FIS/\-04564/\-2019, No. UID/\-04564/\-2020, and No. POCI-01-0145-FEDER-029912 with financial support from Science, Technology and Innovation, 
in its FEDER component, and by the FCT/MCTES budget through national funds (OE).

\end{document}